\documentclass[9pt,twocolumn,twoside]{osajnl}

\journal{ol} 

\setboolean{shortarticle}{true}

\usepackage{lineno}
\linenumbers

\newcommand{\rev}[1]{{\color{black} #1}}

\title{Non-Hermitian Bloch-Zener phase transition}

\author[1,2,*]{Stefano Longhi}

\affil[1]{Dipartimento di Fisica, Politecnico di Milano, Piazza L. da Vinci 32, I-20133 Milano, Italy}
\affil[2]{IFISC (UIB-CSIC), Instituto de Fisica Interdisciplinar y Sistemas Complejos - Palma de Mallorca, Spain}
\affil[*]{Corresponding author: stefano.longhi@polimi.it}

\begin{abstract}
Bloch-Zener oscillations (BZO), i.e. the interplay between Bloch oscillations and Zener tunneling in two-band lattices under an external dc force, are ubiquitous in different areas of wave physics, including photonics. While in Hermitian systems such oscillations are rather generally aperiodic and only accidentally periodic, \rev{in non-Hermitian (NH) lattices BZO can show a transition from aperiodic to periodic as a NH parameter in the system is varied. Remarkably, the phase transition can be either smooth or sharp, contrary to other types of NH phase transitions which are universally sharp. A discrete-time photonic quantum walk on a synthetic lattice is suggested for an experimental observation of smooth BZO phase transitions.}
\end{abstract}

\setboolean{displaycopyright}{true}

\begin{document}

\maketitle

%





{\em Introduction.} 
Bloch oscillations and Zener tunneling \cite{r1,r2,r3} are ubiquitous phenomena of coherent quantum and classical wave transport in lattice systems under the action of a dc force. Such phenomena have been observed in a variety of physical systems, ranging from solid-state superlattices to matter-wave systems, acoustic and photonic lattices. While in multiband lattices Bloch oscillations are usually damped because of wave tunneling into higher-order lattice bands (Zener tunneling) \cite{r4,r5},
in a two-band lattice the interplay between Bloch oscillations and Zener tunneling leads to a generally aperiodic type of undamped oscillations, referred to as Bloch-Zener oscillations (BZO)  \cite{r6,r7}. BZO have been theoretically studied and experimentally observed in a wide variety of photonic structures \cite{r6,r7,r8,r9,r10,r11,r12,r13,r14,r15,r16}. In such systems, the energy spectrum consists of two displaced Wannier-Stark (WS) ladders, which explains the aperiodic feature of BZO and the very accidental occurrence of a periodic dynamics under special tuning of parameters.\\
 Recently, great attention has been devoted to study the spectral, topological and transport properties of systems described by effective non-Hermitian (NH) Hamiltonians \cite{r17}, and the phenomenon of BZO has been extended to NH lattices \cite{r18,r19,r20,r21,r22,r23,r24,r25} with the prediction of WS ladder exceptional points and chiral Zener tunneling \cite{r19,r20,r23,r25} as unique features of NH dynamics. Contrary to Hermitian systems, where BZO are aperiodic and only accidentally periodic, in NH systems the two WS ladders can display two different decay (or growth) rates, resulting in a periodic dynamics after an initial transient because of the dominance of one set of WS ladder eigenmodes over the other one.  Interestingly, when a NH parameter in the system is varied, a phase transition can arise separating a region of real WS energies, corresponding to aperiodic BZO, to complex WS energies, corresponding to periodic BZO. Such a kind of NH phase transition and its dynamical implications have been little studied and have remained so far elusive in experiments. Discrete-time photonic quantum walks (QW) in coupled-fiber loops 
\cite{r26} have provided in the past few years a fantastic platform for the observation of NH phase transitions using synthetic lattices \cite{r27,r28,r29,r30}, such as parity-time ($\mathcal{PT}$) \cite{r27}, localization-delocalization \cite{r29} and  topological \cite{r30} phase transitions.\\
In this Letter we predict \rev{ the existence of BZO phase transitions, from aperiodic to periodic oscillations, when a NH  parameter in the system is increased. The phase transition is related to a spectral phase transition of the WS quasi-energy ladders and, contrary to typical $\mathcal{PT}$ symmetry breaking phase transitions, it can appear either sharp or smooth.  A specially-designed photonic QW is suggested as an experimentally-accessible platform for the observation of BZO phase transitions.}\\

 \begin{figure}
  \centering
    \includegraphics[width=0.45\textwidth]{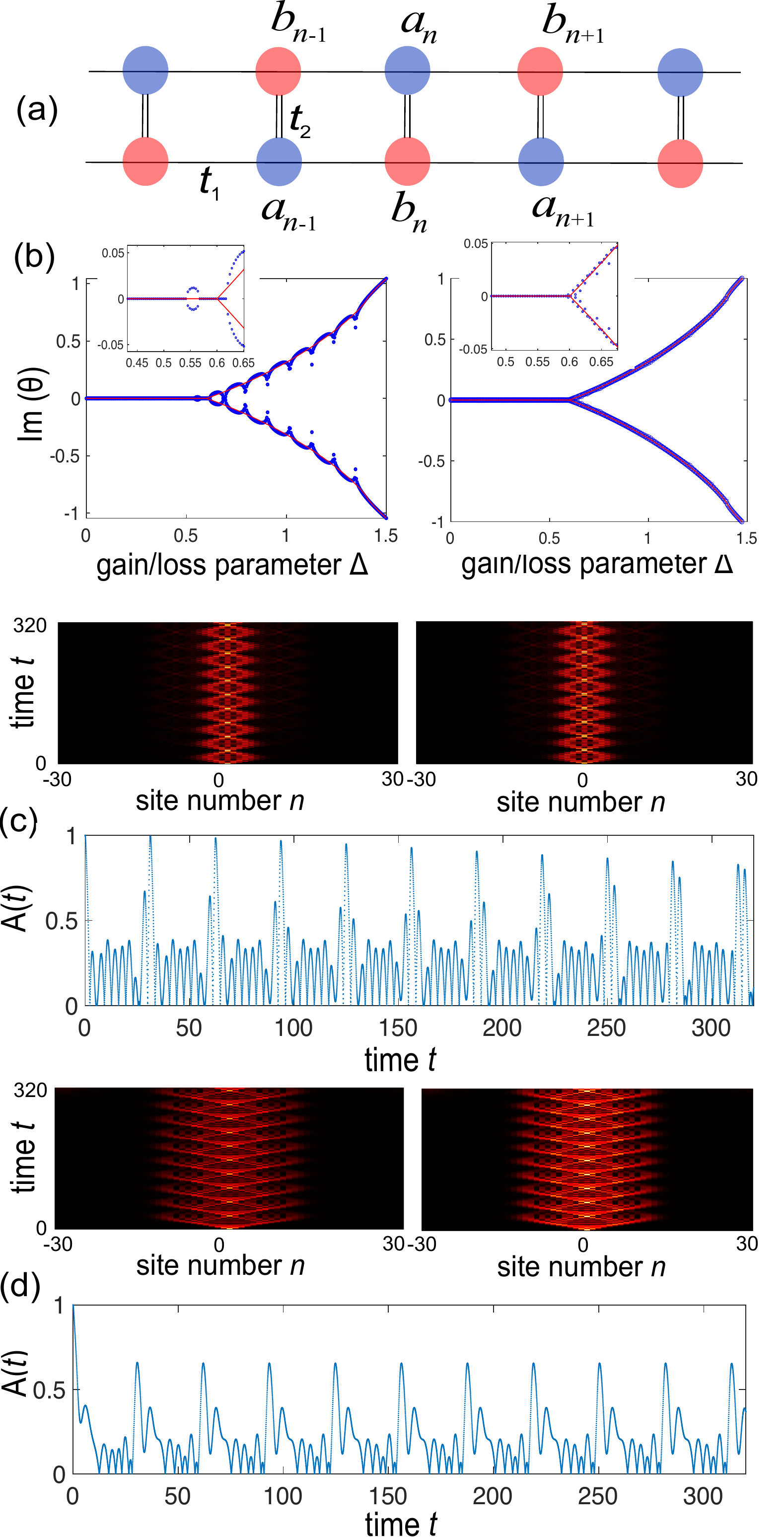}
    \caption{(a) Schematic of the binary NH lattice. Complex on-site potential $\pm i \Delta$, corresponding to gain/loss terms, are applied to sublattices A (blue circles) and B (red circles), respectively. (b) Numerically-computed behavior (dots) of the imaginary part of $\theta$, defining the growth/decay rate of the WS ladders, for increasing values of $\Delta$. Hopping amplitudes are $t_1=0.2$ and $t_2=1$. The dc force is $F=0.2$ in the left panel, and $F=0.02$ in the right panel. The solid red curve, almost overlapped with the dots, is the prediction of ${\rm Im}(\theta)$ provided by the WKB method. The insets depict an enlargement of the curves near the critical value $\Delta_c=t_1-2t_2=0.6$, clearly showing that the phase transition is sharp. (c) Temporal evolution of the normalized amplitudes $|\tilde{a}_n(t)|$ and $|\tilde{b}_n(t)|$ in the two sublattices A and B in the presence of the dc force $F=0.2$ for $\Delta=0.4$ ($\theta$ real). The temporal evolution of the normalized revival amplitude $A(t)=|\tilde{a}_0(t)|$ is also depicted, displaying aperiodic dynamics. (d) Same as (c), but for $\Delta=0.7$ ($\theta$ complex). Note that, after an initial transient, the dynamics becomes periodic with period $T_1= 2 \pi /F$.}
\end{figure}

{\it Bloch-Zener phase transitions in continuous-time NH lattices.}  To illustrate the BZO phase transition, let us first consider a rather general two-band tight-binding lattice under continuous-time evolution, and let $a_n(t)$, $b_n(t)$ be the wave amplitudes in the two sublattices A and B of the $n$-th unit cell. The external dc force $F$ introduces an on-site potential $Fn$ that increases linearly with the cell number $n$. Let us indicate by $\mathcal{H}(k)$ the $2 \times 2$ matrix Hamiltonian describing the lattice in Bloch space in the absence of the dc force, where $- \pi \leq k < \pi$ is the Bloch wave number, and let us assume that $\mathcal{H}_{2,2}(k)=-\mathcal{H}_{1,1}(k)$, so that energies appear in pairs $(E,-E)$. The dispersion curves of the two Bloch bands read $E_{\pm}(k)= \pm \sqrt{\mathcal{H}_{1,1}^2(k)+\mathcal{H}_{1,2}(k)\mathcal{H}_{2,1}(k)}$. When the dc force $F$ is applied to the system, the energy spectrum is pure point and composed by two interleaved WS ladders, namely one has $E_{\pm}=lF \pm \theta$ where $l=0, \pm1, \pm2,....$, $ \theta=F \varphi/(2 \pi)$ and $\varphi$ is a generally complex angle such that $\cos \varphi$ 
is the half trace of the $2 \times 2$ ordered exponential matrix \cite{r23}
\begin{equation}
U= \int_{-\pi}^{\pi} dk \exp \{-i \mathcal{H} (k)/F   \},
\end{equation}
i.e. $ \cos \varphi=(U_{1,1}+U_{2,2})/2$.
In an Hermitian lattice, the angle $\varphi$ is real and the dynamics in the time domain is generally aperiodic and 
characterized by two time periods \cite{r6}. The first one, $T_1=2 \pi/F$,
is determined by the mode spacing of each WS ladder, whereas the second
one, $T_2= (\pi / \varphi) T_1$, is
determined by the shift of the two interleaved WS ladders. Note that only accidentally, when $\varphi/  \pi$ is a rational number, or in the trivial.strong forcing regime the dynamics is periodic. In a NH lattice, the angle $\varphi$ can be either real or complex. In the former case, like in the Hermitian case the dynamics is characterized by two periods and is rather generally aperiodic. Conversely, in the second case the imaginary part of $\theta$ defines the decay/growth rate of the two WS ladders and after an initial transient the dynamics in the time domain is dominated by one WS ladder and it is thus periodic with period $T_1$. The determination of  $\theta$ should be done rather generally only numerically. {\rev{Approximate analytical results can be obtained in the strong and weak forcing regimes. In the former case each unit cell in the lattice is decoupled from the other ones, the two WS ladders originate from the splitted energies of each dimer in the chain and the dynamics is trivial , displaying Rabi-like oscillations in each dimer below a possible symmetry-breaking point. In the latter case, $F \rightarrow 0$,}  a rough estimate of $\theta$ can be obtained by a standard WKB analysis. Neglecting the NH Berry phase contribution in the adiabatic cycle, the approximate form of $\theta$ reads \cite{r23}
\begin{equation}
\theta \simeq \frac{1}{2 \pi} \int_{-\pi}^{\pi} dk E_+(k)=\frac{1}{2 \pi} \int_{-\pi}^{\pi} dk \sqrt{\mathcal{H}_{1,1}^2(k)+\mathcal{H}_{1,2}(k)\mathcal{H}_{2,1}(k)}.
\end{equation}
\begin{figure}[h]
  \centering
    \includegraphics[width=0.45\textwidth]{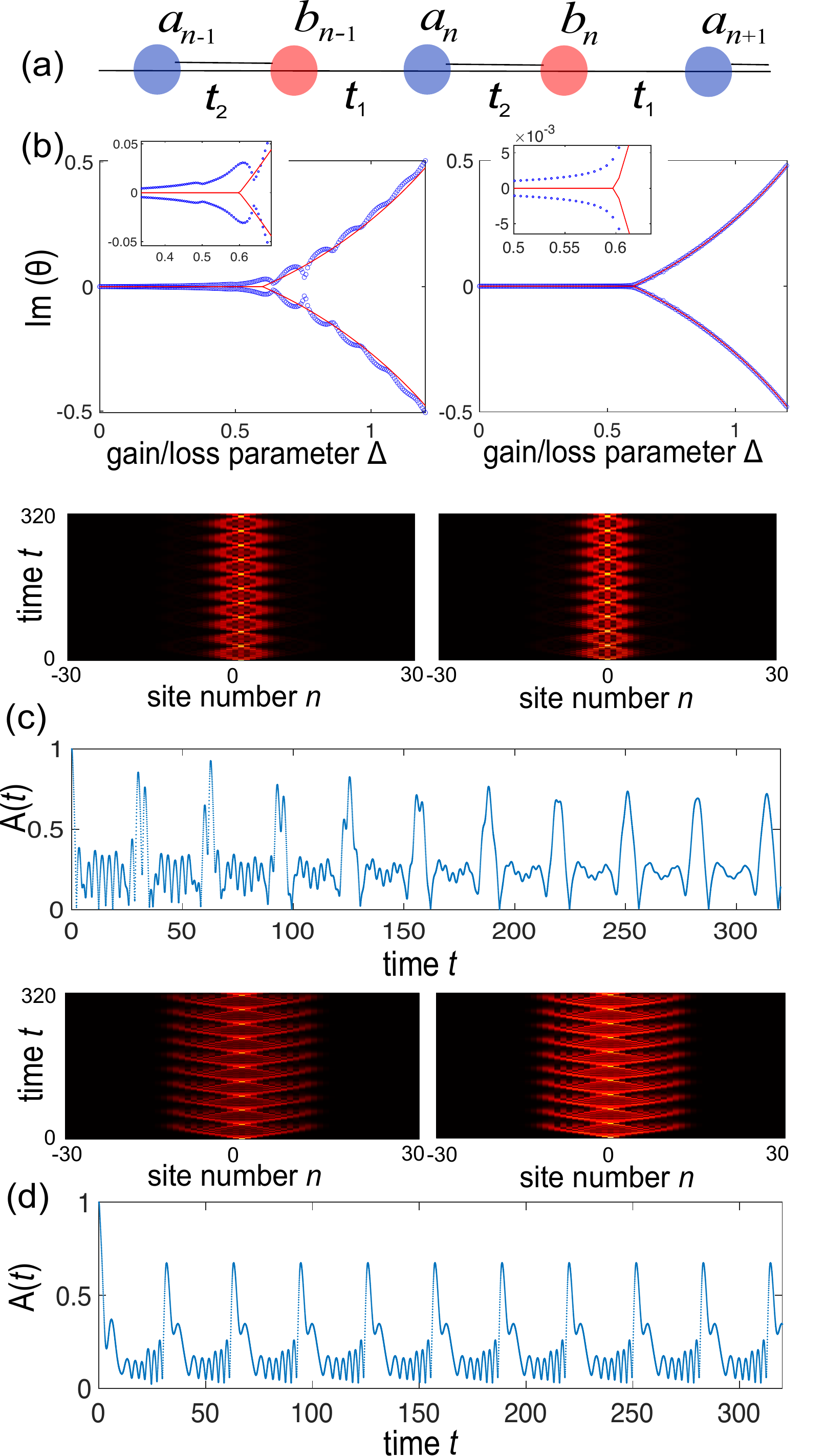}
    \caption{Same as Fig.1, but for the NH Rice-Mele model. Hopping amplitudes are $t_1=0.4$ and $t_2=1$. Other parameter values are as in Fig.1. Note that, unlike the model of Fig.1, the phase transition is smooth, as shown in the insets of panels (b).}
\end{figure}
Interestingly, when a NH parameter in the system is varied, a transition can occur from a real to a complex $\theta$, typically via a coalescence of the two WS ladders and the appearance of a WS exceptional point \cite{r19,r20}. \rev{A remarkable feature of the BZO phase transition, which is rather unique as compared to other types of NH phase transitions, is that it can be exact (sharp) or approximate (smooth)}. We illustrate such an interesting result by considering two NH binary lattice models, shown in Figs.1(a) and 2(a), where non-Hermiticity is introduced by balanced gain/loss terms $ \pm i \Delta$ in the two sublattices A and B.\\ 
The first model, shown in Fig.1(a), was previously introduced in Ref.\cite{r20}. For this model, the elements of the matrix Bloch Hamiltonian read $\mathcal{H}_{1,1}(k)=-\mathcal{H}_{2,2}(k)=i \Delta$ and $\mathcal{H}_{1,2}(k)=\mathcal{H}_{2,1}(k)=t_2+2 t_1 \cos k$, where $t_1$ and $t_2$ are the inter- and intra-dimer hopping amplitudes with $t_2>2 t_1>0$. In the absence of the dc force, the Hamiltonian shows a $\mathcal{PT}$ symmetry breaking phase transition as $\Delta$ is increased above the critical value $\Delta_c=t_2-2t_1$. As the dc force $F$ is applied, a similar phase transition is found for WS ladders, from real $\theta$ to complex $\theta$, as $\Delta$ is increased \cite{r20}. The phase transition is sharp, corresponds to the coalescence of  WS energy spectra ($\theta=0$) with their eigenmodes (WS exceptional point \cite{r20}), and occurs at a critical value close to $\Delta_c$, as predicted by the WKB analysis; see Fig.1(b).\\
 The second model, shown in Fig.2(a), is a NH extension of the Rice-Mele Hamiltonian \cite{r35}, corresponding to $\mathcal{H}_{1,1}(k)=-\mathcal{H}_{2,2}(k)=i \Delta$ and $\mathcal{H}_{1,2}(k)=\mathcal{H}_{2,1}^*(k)=t_2+ t_1 \exp(- ik)$ for the Bloch Hamiltonian. For $F=0$, the system displays a  $\mathcal{PT}$ symmetry breaking phase transition as $\Delta$ is increased above the critical value $\Delta_c=|t_2-t_1|$. When the dc force $F$ is applied, a phase transition is found for WS ladders as $\Delta$ is increased above $\sim \Delta_c$. However, unlike the previous model the phase transition is smooth and the WS exceptional point is avoided: the imaginary part of $\theta$, i.e. amplification/damping rate of the WS ladders, continuously changes from small to large values as $\Delta$ is increased, with a rather abrupt increase at $\Delta \sim \Delta_c$; see Fig.2(b). \rev{The small (but non-vanishing) imaginary part of $\theta$ for $\Delta < \sim \Delta_c$, that makes the phase transition smooth, basically arises from the complex Berry phase of the adiabatically-evolving Hamiltonian eigenstates \cite{r18,referee1,referee2}, which is neglected in Eq.(2)}.\\
 In both models, the WS phase transition corresponds in the time domain to the transition from an aperiodic to a period dynamics, as shown in Figs.1(c,d) and 2(c,d). The figures illustrate the temporal evolution of the normalized amplitudes $\tilde{a}_n(t)=a_n(t) / \sqrt{ \sum_n ( |a_n(t)|^2+b_n(t)|^2) }$ and $\tilde{b}_n(t)=b_n(t) / \sqrt{ \sum_n ( |a_n(t)|^2+b_n(t)|^2) }$ versus time $t$ in the two sublattices for initial excitation condition $a_n(0)=\delta_{n,1}$ and  $b_n(0)=0$. The behavior of the revival amplitude $A(t)=|\tilde{a}_0(t)|$ is also depicted, clearly showing the onset of aperiodic ($\theta \sim$ real) or aperiodic ($\theta$ complex) dynamics over a time window $10 \times T_1$. 

{\it Bloch-Zener phase transition in discrete-time photonic quantum walks.}
We consider a discrete-time photonic QW realized using optical pulses in a synthetic mesh lattice realized by  two coupled fiber loops of slightly different lengths, which has been discussed and experimentally implemented in several previous works (see e.g. \cite{r21,r26,r27,r28,r29,r30} and reference therein). 
To realize a binary lattice model, we consider a two-step process \cite{r28} which is described by the following set of discrete-time coupled-mode equations
 \begin{eqnarray}
 u^{(m-1/2)}_n & = & \left[   \cos \beta_1 u^{(m-1)}_{n+1}+i \sin \beta_1 v^{(m-1)}_{n+1}  \right]  \exp (-i\phi_1^{(m)}) \; \\
 v^{(m-1/2)}_n & = & \left[   \cos \beta_1 v^{(m-1)}_{n-1}+i \sin \beta_1 u^{(m-1)}_{n-1}  \right] \exp (i\phi_1^{(m)}) 
 \end{eqnarray}
in the first step, and
 \begin{eqnarray}
 u^{(m)}_n & = & \left[   \cos \beta_2 u^{(m-1/2)}_{n+1}+i \sin \beta_2 v^{(m-1/2)}_{n+1}  \right]  \exp (-i\phi_2^{(m)}) \; \\
 v^{(m)}_n & = & \left[   \cos \beta_2 v^{(m-1/2)}_{n-1}+i \sin \beta_2 u^{(m-1/2)}_{n-1}  \right] \exp (i\phi_2^{(m)}) 
 \end{eqnarray}
in the second step. In the above equations, $u_n^{(m)}$ and $v_n^{(m)}$ are the pulse amplitudes at discrete time step $m$ and lattice site $n$ in the two fiber loops, $\beta_1$ and $\beta_2$ are the coupling angles and $\phi_1(m)$, $\phi_2(m)$ the complex phases applied in the two steps. 
In order to realize a discrete-time version of the driven NH Rice-Mele model of Fig.2(a), we assume the following phases
\begin{equation}
\phi_{1,2}^{(m)}=\mp i (\Delta /2)+Fm\pm \pi/2  
\end{equation}
where $F$ and $\Delta$ are real parameters and $F= 2 \pi/M$, with $M$ integer.
In fact, as shown in Sec.2 the Supplementary document the continuous-time limit of the two-step QW, obtained by assuming $\beta_{1,2}= \pi/2 \mp t_{2,1}$, with $t_{1,2}$, $\Delta$ and $F$ small numbers, reproduces  the NH Rice-Mele model of Fig.2(a) with hopping amplitudes $t_1$, $t_2$, dc force $2F$ and gain/loss parameter $\Delta$.\\
Owing to the discrete translation invariance of the lattice, we may set $u_n^{(m)}=x_m \exp(iqn)$, $v_n^{(m)}=y_m \exp(iqn)$, where $-\pi \leq q < \pi$ is the Bloch wave number and the amplitudes $x_m$, $y_m$ evolve according to the map $(x_m,y_m)^T=U^{(m)} ( x_{m-1}, y_{m-1})^T$,
where the elements of the $2 \times 2$ matrix $U^{(m)}$ are given in Sec.1 of the Supplemental document. In the absence of the dc force, $F=0$, the elements of the $U$-matrix do not depend on time step $m$, and the eigenenergies $ E_{\pm}(q)$ of the lattice are just the Floquet exponents of the matrix $U$ and read
\begin{equation}
E_{\pm}(q)= \pm {\rm {acos} } \left\{ \cos \beta_1 \cos \beta_2 \cos(2q)+ \sin \beta_1 \sin \beta_2 \cosh \Delta  \right\}
\end{equation}

which define the dispersion curves of the two lattice bands.
Note that a spectral phase transition, from $E_
{\pm}(q)$ real to complex, is observed as the gain/loss parameter $\Delta$ is increased above the critical value $\Delta_c$ such that 
$ | \sin \beta_1 \sin \beta_2 | {\rm cosh}  \Delta_c=1-| \cos \beta_1 \cos \beta_2 |$.\\ When the dc force $F$ is applied, the quasi-energy spectrum is given by $E_{\pm}(q)= \pm (1/M) \mu(q) \equiv \pm \theta(q)$, where $ \pm \mu(q)$ are the Floquet exponents of the time-ordered propagator $S=U^{(M)} \times U^{(M-1)} \times... \times U^{(1)}$. Numerical computation of the Floquet exponents of the propagator $S$ (see Supplemental document) indicate that $\theta(q)$ does not depend on $q$, i.e. the quasi-energy bands are flat: in other words, the dc force induce a band collapse. The band collapse basically corresponds to the appearance of two interleaved WS ladders, like in the continuous-time QW models previously studied. As the gain/loss parameter $\Delta$ is increased above zero, like for the NH Rice-Mele model one observes a smooth phase transition, from almost real $\theta$ to complex $\theta$, as shown in Fig.3(a). In the time-domain, the spectral phase transition corresponds to a change of the dynamics, as $\Delta$ is increased, from aperiodic ($\theta \sim$ real) to periodic ($\theta$ complex), after an initial transient. This transition is shown in Figs. 3(b) and (c) for typical values of parameters that are feasible for an experimental implementation.

\begin{figure}[H]
  \centering
    \includegraphics[width=0.45\textwidth]{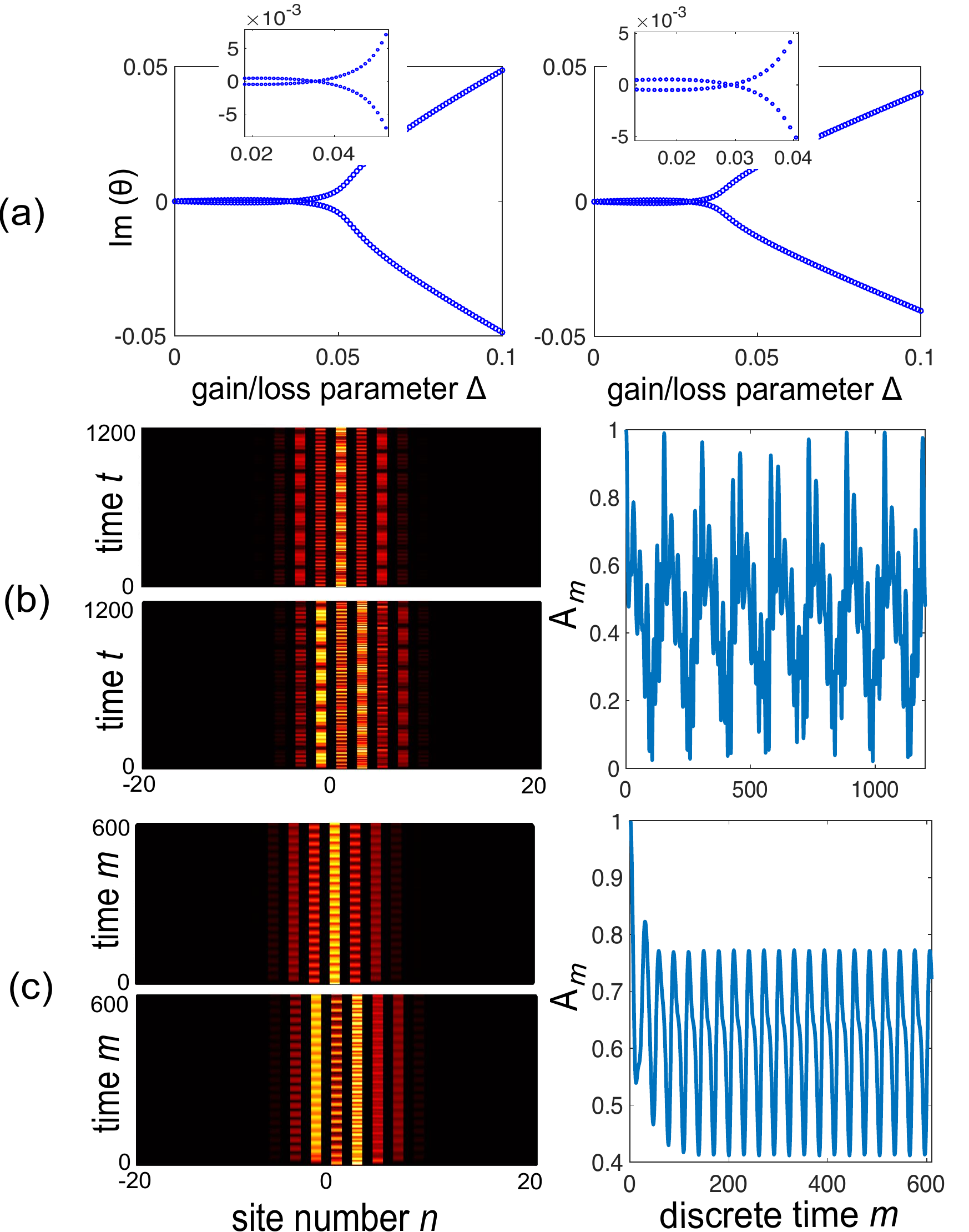}
    \caption{NH Bloch-Zener phase transition in a discrete-time photonic QW. (a) Behavior of the imaginary part of $\theta$ versus gain/loss parameter $\Delta$ for a dc force $F= 2 \pi /M$ with $M=61$ (left panel) and $M=102$ (right panel). The coupling angles are $\beta_1= \pi/2-0.1$ and $\beta_2= \pi/2-0.15$. Insets show an enlargement of the curves near the phase transition point. (b,c) Temporal dynamics in the coupled-fiber loops for single-pulse excitation of the system ($u_n^{(0)}=\delta_{n,0}$, $v_n^{(0)}=0$) and for a dc force $F=2 \pi /M$ with $M=61$. The left panels show the discrete time evolution of the normalized pulse amplitudes $\tilde{u}_n^{(m)}=u_n^{(m)}/ \sqrt{|u_n^{(m)}|^2+|v_n^{(m)}|^2}$ (upper plot) and $\tilde{v}_n^{(m)}=v_n^{(m)}/ \sqrt{|u_n^{(m)}|^2+|v_n^{(m)}|^2}$ (lower plot). The right panels show the temporal evolution of the recurrence amplitude $A_m=|\tilde{u}_0^{(m)}|$. In (b) $\Delta=0.036$, corresponding to $\theta \sim$ real and aperiodic temporal dynamics; in (c) $\Delta=0.06$, corresponding to a complex $\theta$ and periodic temporal dynamics after an initial transient.}
\end{figure}

 {\it Conclusion.} We predicted the existence of phase transitions for Bloch-Zener oscillations in NH lattices under a dc force. The phase transition, observed as a NH parameter in the system is varied, corresponds to a change from aperiodic to periodic oscillations associated to a spectral phase transition of the WS ladders. \rev{A rather unique feature of BZO phase transition, which does not find any analogue in other types of NH phase transitions, is that  the transition can be exact (sharp) or approximate (smooth)}. A specially-designed NH discrete-time photonic QW has been suggested for the observation of BZO phase transitions. \rev{Our results provide major insights into NH phase transitions and could be of relevance to a wide variety of Wannier-Stark NH systems beyond photonics, realized for example in ultracold atoms or in solid-state superconducting systems.}\\
\\
\noindent
{\bf Disclosures}. The author declares no conflicts of interest.\\
\\
{\bf Data availability}. No data were generated or analyzed in the presented research.\\
\\
{\bf Supplemental document}. See Supplement 1 for supporting content.

\newpage


 {\bf References with full titles}\\
 \\
 \noindent

1. F. Bloch,  \"uber die Quantenmechanik der Elektronen in Kristallgittern, Z. Phys. {\bf 52}, 555 (1929).\\
2. C. Zener, A Theory of the Electrical Breakdown of Solid Dielectrics, Proc. R. Soc. A {\bf 145}, 523 (1934). \\
3. G.H. Wannier,
Wave Functions and Effective Hamiltonian for Bloch Electrons in an Electric Field,
Phys. Rev. {\bf 117}, 432 (1960).\\
4. M. Ghulinyan, C.J. Oton, Z. Gaburro, L. Pavesi, C. Toninelli, and D.S. Wiersma, Zener Tunneling of Light Waves in an Optical Superlattice,
Phys. Rev. Lett. {\bf 94}, 127401 (2005).\\
5. H. Trompeter, W. Krolikowski, D.N. Neshev, A.S. Desyatnikov, A.A. Sukhorukov, Y.S. Kivshar, T. Pertsch, U. Peschel, and F. Lederer,
 Bloch Oscillations and Zener Tunneling in Two-Dimensional Photonic Lattices,
Phys. Rev. Lett. {\bf 96}, 053903 (2006).\\
6. B. M. Breid, D. Witthaut, and H.J. Korsch, Bloch-Zener oscillations, New J.Phys. {\bf 8}, 110 (2006).\\
7. S. Longhi, Optical Zener-Bloch oscillations in binary waveguide arrays, EuroPhys. Lett. {\bf 76}, 416 (2006).\\
8. S. Longhi, Bloch oscillations and Zener tunneling with nonclassical light, Phys. Rev. Lett. {\bf 101}, 193902, (2008).\\
9. F. Dreisow, A. Szameit, M. Heinrich, T. Pertsch, S. Nolte, A. T\"unnermann, and S. Longhi, Bloch-Zener Oscillations in Binary Superlattices, Phys. Rev. Lett. {\bf 102}, 076802 (2009).\\
10. S. Kling, T. Salger, C. Grossert, and M. Weitz,
Atomic Bloch-Zener Oscillations and St\"uckelberg Interferometry in Optical Lattices,
Phys. Rev. Lett. {\bf 105}, 215301 (2010).\\
11. S. Longhi,
Bloch-Zener oscillations of strongly correlated electrons,
Phys. Rev. B {\bf 86}, 075144 (2012).\\
12. R. Khomeriki and S. Flach,
Landau-Zener Bloch Oscillations with Perturbed Flat Bands,
Phys. Rev. Lett. {\bf 116}, 245301 (2016).\\
13. Y. Zhang, D. Zhang, Z. Zhang, C. Li, Y. Zhang, F. Li, Mi. R. Belic, and M. Xiao,
Optical Bloch oscillation and Zener tunneling
in an atomic system, Optica {\bf 4}, 571 (2017).\\
14. Y. Sun, D. Leykam, S. Nenni, D. Song, H. Chen, Y.?D. Chong, and Z. Chen,
Observation of Valley Landau-Zener-Bloch Oscillations and Pseudospin Imbalance in Photonic Graphene,
Phys. Rev. Lett. {\bf 121}, 033904 (2018).\\
15. A D'Errico, R Barboza, R Tudor, A Dauphin, P Massignan, L Marrucci, and F. Cardano,
Bloch-Landau-Zener dynamics induced by a synthetic field in a photonic quantum walk,
APL Photonics {\bf 6}, 020802 (2021).\\
16. Z. Zhang, S. Ning, H. Zhong, M. R. Belic, Y. Zhang, Y. Feng, S. Liang, Y. Zhang, and M Xiao,
Experimental demonstration of optical Bloch oscillation in electromagnetically induced photonic lattices,
Fundamental Research {\bf 2}, 401 (2022).\\
17. Y. Ashida, Z. Gong, and M. Ueda, Non-Hermitian physics, Adv. Phys. {\bf 69}, 249 (2020).\\
18. S. Longhi, Bloch oscillations in complex crystals with $\mathcal{PT}$-symmetry, Phys. Rev. Lett. {\bf 103}, 123601 (2009). \\
19 C. Elsen, K. Rapedius, D. Witthaut, and H.J. Korsch, Exceptional points in bichromatic Wannier-Stark systems, J. Phys. B {\bf 44},  225301 (2011).\\
20. N. Bender, H. Li, F. M. Ellis, and T. Kottos, Wave-packet self-imaging and giant recombinations via stable Bloch-Zener
oscillations in photonic lattices with local $\mathcal{PT}$ symmetry,
Phys. Rev. A {\bf 92}, 041803 (2015).\\
21. M. Wimmer, M.-A. Miri, D. Christodoulides, and U. Peschel, Observation of Bloch oscillations in complex $\mathcal{PT}$-symmetric photonic
lattices, Sci. Rep. {\bf 5}, 17760 (2015).\\
22. Y.-L. Xu, W.S. Fegadolli, L. Gan, M.-H. Lu, X.-P. Liu, Z.-Y. Li, A. Scherer, and Y.-F. Chen, 
 Experimental realization of Bloch oscillations in a parity-time synthetic silicon photonic lattice, Nat. Commun. {\bf 7}, 11319 (2016).\\
23. S. Longhi, Non-Bloch-Band Collapse and Chiral Zener Tunneling, Phys. Rev. Lett. {\bf 124}, 066602 (2020).\\
24. S. Xia, C. Danieli, Y. Zhang, X. Zhao, H. Lu, L. Tang, D. Li, D. Song,and Z. Chen,
Higher-order exceptional point and Landau-Zener Bloch oscillations in driven non-Hermitian photonic Lieb lattices,
APL Photonics {\bf 6}, 126106 (2021).\\
25. L. Zheng, B. Wang, C. Qin, L. Zhao, S. Chen, W. Liu, and P. Lu,
Chiral Zener tunneling in non-Hermitian frequency lattices, Opt. Lett. {\bf 47}, 4644 (2022).\\
26.  A. Schreiber, K. N. Cassemiro, V. Potocek, A. Gabris, P. J. Mosley, E. Andersson, I. Jex, and
C. Silberhorn,  Photons Walking the Line: A Quantum Walk with Adjustable Coin Operations, Phys. Rev. Lett. {\bf 104}, 050502 (2010).\\
27. A. Regensburger, C. Bersch, M. A. Miri, G. Onishchukov, D.N. Christodoulides, and U. Peschel, Parity-time synthetic photonic lattices,
Nature {\bf 488}, 167 (2012).\\
28. S. Weidemann, M. Kremer, T. Helbig, T. Hofmann, A. Stegmaier, M. Greiter, R. Thomale, and
A. Szameit, Topological funneling of light, Science {\bf 368}, 311 (2020).\\
29. S. Weidemann, M. Kremer, S. Longhi, and A. Szameit, Coexistence of dynamical delocalization and spectral localization through stochastic dissipation, Nature Photon. {\bf 15}, 576 (2021).\\
30. S. Weidemann, M. Kremer, S. Longhi, and A. Szameit, Topological triple phase transition in non-Hermitian Floquet quasicrystals, Nature {\bf 601}, 354 (2022).\\
31. \rev{H Zhao, S Longhi, and L Feng, Robust Light State by Quantum Phase Transition in Non-Hermitian Optical Materials,
Sci. Rep. {\bf 5}, 17022 (2015).}\\
32. \rev{X.Z. Zhang and Z. Song, Partial topological Zak phase and dynamical confinement in a non-Hermitian bipartite system, Phys. Rev. B {\bf 99}, 012113 (2021).}\\
33. M. Demirplak and S. A. Rice, Adiabatic Population Transfer with Control Fields, J. Phys. Chem. A {\bf 107}, 9937
(2003).

\end{document}